\documentclass[10pt,onecolumn,oneside]{article}
\usepackage{amsmath,amssymb,amsfonts,amsthm,dsfont}
\usepackage{authblk}
\usepackage{verbatim}
\usepackage{longtable}
\usepackage{xcolor}
\usepackage{multicol}

\allowdisplaybreaks 

\theoremstyle{plain} \newtheorem{lemma}{\textbf{Lemma}}
\theoremstyle{plain} \newtheorem{proposition}{\textbf{Proposition}}
\theoremstyle{remark} \newtheorem{remark}{\textbf{Remark}}
\theoremstyle{plain} \newtheorem{theorem}{\textbf{Theorem}}
\theoremstyle{plain} \newtheorem{example}{\textbf{Example}}
\theoremstyle{plain} \newtheorem{corollary}{\textbf{Corollary}}
\theoremstyle{definition} \newtheorem{definition}{\textbf{Definition}}
\theoremstyle{plain}

\newtheoremstyle{thmstyleAAA3}{3pt}{3pt}{}{}{\bfseries}{}{0.5em}{}
\theoremstyle{thmstyleAAA3} \newtheorem{cumjacere}{\textbf{Conjecture}}
\newcommand{\rmv}[1]{}

\makeatletter
\let\@@pmod\pmod
\DeclareRobustCommand{\pmod}{\@ifstar\@pmods\@@pmod}
\def\@pmods#1{\mkern4mu({\operator@font mod}\mkern 6mu#1)}
\makeatother

\begin{document}

\title{Unicyclic strong permutations}
\author[*]{Claude Gravel}
\author[**]{Daniel Panario}
\author[**]{David Thomson}
\affil[*]{\small{Unaffiliated Researcher, \mbox{claudegravel1980@gmail.com}}}
\affil[**]{\small{School of Mathematics and Statistics, Carleton University, Canada, \mbox{\{daniel,dthomson\}@math.carleton.ca}}}
\date{\today}
\maketitle

\begin{abstract}
In this paper, we study some properties of a certain kind of permutation $\sigma$ over $\mathbb{F}_{2}^{n}$, where $n$ is a positive integer.
%
%
The desired properties for $\sigma$ are: (1) the algebraic degree of each component function is $n-1$; (2) the permutation is unicyclic; (3) the number of terms of the algebraic normal form of each component is at least $2^{n-1}$. We call permutations that satisfy these three properties simultaneously unicyclic strong permutations. We prove that our permutations $\sigma$ always have high algebraic degree and that the average number of terms of each component function tends to $2^{n-1}$. We also give a condition on the cycle structure of $\sigma$. We observe empirically that for $n$ even, our construction does not provide unicylic permutations. For $n$ odd, $n \leq 11$, we conduct an exhaustive search of all $\sigma$ given our construction for specific examples of unicylic strong permutations. We also present some empirical results on the difference tables and linear approximation tables of $\sigma$.
%
%


\end{abstract}
\section{Notation, facts and definitions}\label{sect:intro}
Let $n$ be a positive integer. Denote by $\mathbb{F}_2$ and $\mathbb{F}_{2^n}$ the finite fields of $2$ and $2^n$ elements, respectively, and denote by $\mathbb{F}_2^n$ the vector space of dimension $n$ over $\mathbb{F}_2$.

Let $N = \{0, 1, \ldots, n-1\},$ let $a = (a_0, \ldots, a_{n-1})$ and for $I \subseteq N$ denote by $a_I = \prod_{i \in I} a_i$. The \emph{algebraic normal form} of the Boolean function $\varphi(a)$ on the $n$ variables $a = (a_0, \ldots, a_{n-1})$ is the unique representation of $f$ in $\mathbb{F}_{2}[a_0, \ldots, a_{n-1}]/(a_0^2 -a_0, \ldots, a_{n-1}^2-a_{n-1})$ of the form
\[ \varphi(a) = \sum_{I \in \mathcal{P}(N)}  x_I  a_I , \quad x_I \in \{0,1\}, \]
where $\mathcal{P}(N)$ denotes the power set of $N$. The \emph{algebraic degree} of $\varphi$ is the minimum degree of the nonzero terms of the ANF of $\varphi$. Moreover, the algebraic degree of any vectorial Boolean function $\varphi(a) = (\varphi_0(a), \ldots, \varphi_{n-1}(a))$ is the maximum algebraic degree of the  component functions $\varphi_0, \ldots, \varphi_{n-1}$.

%



Let $\mathcal{I}_n$ be the set of irreducible polynomials of degree $n$ over $\mathbb{F}_2$ and let $Q\in\mathcal{I}_n$.
For $a=(a_0,\ldots,a_{n-1})\in \mathbb{F}_{2}^{n}$, we denote by $P_a$ the polynomial associated canonically to $a$
\begin{displaymath}
P_{a}(X)=a_{0}+a_{1}X+\cdots+a_{n-1}X^{n-1}\in\mathbb{F}_{2}[X]\slash (Q).
\end{displaymath}
%
For $a\in \mathbb{F}_{2}^{n}$, $t\in [1, 2^n-2]$ and $0\leq j \leq n-1$, we denote by $\varphi_{j,t}$ the boolean functions such that
\begin{equation}
\left(P_{a}(X)\right)^{t}\equiv \sum_{j=0}^{n-1}{\varphi_{j,t}(a)X^{j}}\pmod*{Q}.\label{powermap}
\end{equation}
We define a permutation $\sigma$ on $\mathbb{F}_2^n$ as the concatenation of several maps
\begin{equation} \begin{array}{clll}
\sigma \colon &\mathbb{F}_2^n &\longrightarrow \mathbb{F}_2[X]/(Q) &\longrightarrow \mathbb{F}_2[X]/(Q) \\
              &&&\longrightarrow \mathbb{F}_2^n \\
              & a &\longmapsto P_a(X)\pmod*{Q} &\longmapsto \left(P_a(X)\right)^t\pmod*{Q}  \\
							&&&\longmapsto \left(\varphi_{0,t}(a), \ldots, \varphi_{n-1,t}(a)\right).
\end{array} \label{eqn:sigma-noperturb} \end{equation}
It is necessary and sufficient that
$\gcd(t,2^n-1) = 1$ for $\sigma$ to define a permutation on
$\mathbb{F}_2^n$, so that the $t$-th power map defines a
permutation of $\mathbb{F}_{2^n}$.
For the remainder of this section, we assume that $\gcd(t, 2^n-1) = 1$ and drop the subscript $t$ from $\varphi_{j,t}$.





We are interested in several properties of boolean functions. We are primarily concerned with boolean functions that have high algebraic degree
for resilience against algebraic attacks
and that have a large number
of terms in their ANF~\cite{Carlet1,Carlet2,MullenPanario2013}.
When the vector of boolean functions $\varphi_j$ represents a
permutation, entries of the difference and linear approximation tables
\cite{Carlet1,Carlet2,XiaoM88,Sigenthaler1984} of it, the cycle structure and the period
\cite{Flajolet90randommapping,flajolet2009analytic,Szpankowski2001},
the number of bits required to describe, or generate, it
\cite{BaBoHKTS2017,Szpankowski2001} and the possibility to
generate it on the fly \cite{BrKa1988} are also interesting
properties. As permutations, they should be comprised of a
single cycle for resilience against potential Fourier-like
attacks in
non-commutative symmetric groups.
To be considered strong for cryptographic purposes,
candidate functions should also exhibit resilience against
linear~\cite{Nyberg2017} and differential
cryptanalysis~\cite{matsui1993linear}.

\begin{definition}
A permutation $\sigma$ on $n$ bits is a \emph{unicyclic strong permutation} if it satisfies the following $3$ properties:
\begin{enumerate}
\item[$\mathfrak{P}$1.] the algebraic degree of $\varphi_{j}$ is $n-1$ for all $0\leq j \leq n-1$;
\item[$\mathfrak{P}$2.] $\sigma$ has one cycle of length $2^{n}$; and
\item[$\mathfrak{P}$3.] the number of terms in the expressions $\varphi_j$ is at least $2^{n-1}$ for all $j$.
\end{enumerate}\label{def:usp}
\end{definition}

Our goal in this work is to find permutations that simultaneously satisfy the properties $\mathfrak{P}1$, $\mathfrak{P}2$ and $\mathfrak{P}3$ from Definition~\ref{def:usp}; that is, our goal is to construct unicyclic strong permutations.
Property $\mathfrak{P}$1 (algebraic degree) is important to prevent algebraic attacks.
Property $\mathfrak{P}$2 (unicyclicity) is to avoid short cycles in general and to avoid decomposability of attacks onto smaller permutations induced by the cycle structure.
Property $\mathfrak{P}$3 (number of terms) is important to prevent algebraic attacks like linearization, exploiting hidden structures, optimization/SAT based attacks and so on.
We also observe that to be considered strong for cryptographic purposes, candidate functions should exhibit resilience against linear~\cite{matsui1993linear} and differential cryptanalysis~\cite{Biham1991}.

We present our construction of unicyclic strong permutations in Section~\ref{app:sect:expansion}, and in Section \ref{sect:specpoly}, we give results on Properties $\mathfrak{P}$1, $\mathfrak{P}$2, and $\mathfrak{P}$3 for our construction. In Section~\ref{sect:compo} we give some empirical results for our permutations. We observe experimentally that our construction only provides unicyclic permutations for $n$ odd; so far we are unaware of a theoretical justification excluding the cases when $n$ is even. Although our focus in this paper is to introduce permutations satisfying Properties~$\mathfrak{P}1$, $\mathfrak{P}2$ and $\mathfrak{P}3$, as already mentioned, before they can be suggested for practical use they must be found to have good resilience against linear and differential cryptanalysis. Some empirical results on the \emph{linearity} and \emph{differential uniformity} of such functions are also presented in Section~\ref{sect:compo}.


\section{Strong permutations}\label{app:sect:expansion}
In this section, we define a permutation $\sigma$ as the composition of $n$ permutations $\sigma_k$, $k=0, 1, \ldots, n-1$. Although we are not formally defining a cipher, we find it helpful to think of $\sigma_k$ as \emph{round functions}, for example as the rounds of a substitution-permutation network. Our precise selection of $\sigma_k$ is motivated both by perturbations and as generalizations of the inverse function used in the S-boxes of the Advanced Encryption Standard (AES)~\cite{daemen2013design}. We define and present some basic facts about the permutations $\sigma_k$. In Section~\ref{sect:specpoly}, we analyze local properties of $\sigma_k$ to derive results on Properties~$\mathfrak{P}1$, $\mathfrak{P}2$ and $\mathfrak{P}3$ for $\sigma$.

Throughout the remainder of this paper, since $2^n-2^k-1\equiv -2^{k}\pmod*{2^{n}-1}$, we often use $-2^k$ to denote this power. We also assume that all calculations in the finite field defined with modulus $Q$ are given by canonical residues after reduction modulo $Q$. In calculations we often suppress the $\pmod*{Q}$.

\subsection{Permutations from perturbations}\label{sec:permsfromperbs}

For a positive integer $n$, let $Q\in\mathcal{I}_{n}$ and let $b\in \mathbb{F}_{2}^{n}$ with $P_{b}(X)\in\mathbb{F}_{2}[X]$ be its canonical representation. Our main idea is the construction of a permutation $\sigma$ as a composition of permutations
\begin{equation}
\sigma = \sigma_{n-1}\cdots \sigma_1\sigma_0
\label{eqn:sigma}
\end{equation} of a special form. Each $\sigma_k$ is a modified power map, as in Equation~\eqref{eqn:sigma-noperturb}: for $k = 0, 1, \ldots, n-1$, we define $\sigma_k$ as
\[
\begin{array}{clll}
\sigma_k\colon &\mathbb{F}_2^n &\longrightarrow \mathbb{F}_2[X]/(Q) & \longrightarrow \mathbb{F}_2[X]/(Q) \\
               &&& \longrightarrow \mathbb{F}_2^n,\\
               & a & \longmapsto P_a(X)\pmod*{Q} &\longmapsto (P_a(X) + P_b(X))^{t_k}\pmod*{Q}\\
							 &&& \longmapsto c,
\end{array}
\]
where $t_k \in [1, 2^n-2]$ and $c = \sigma_k(a)$ is the coefficient vector of the residue of $\left(P_a(X) + P_b(X)\right)^{t_k}\pmod*{Q}$.

The novelty in $\sigma_k$ is the addition of a fixed polynomial $P_b\in \mathbb{F}_{2}[X]\slash(Q)$,
that we call a \emph{perturbation polynomial}, to the input of a power map as in Equation~\eqref{powermap}. In effect, this performs a fixed bit flip for the inputs to the power maps. If $n=8$, $Q(X)=X^8+X^4+X^3+X+1$ (Rijndael's polynomial) and $t\equiv -1\pmod*{255}$, with no perturbation under
repeated $t$-th powers we obtain only cycles of length $2$.
However, we find experimentally that for exactly half of all
perturbations $b \in \mathbb{F}_2^n$ the permutation
\[\left(P_a(X) + P_b(X)\right)^{-1}\]
is unicyclic. We discuss more experimental results in
Section~\ref{sect:compo}.



\subsection{Expanded form of a strong permutation}

In this section, we prove that for all perturbation vectors $b \in \mathbb{F}_2^n$ and for all $Q\in \mathcal{I}_n$, the expansion of $\big(P_{a}(X)+P_{b}(X)\big)^{-2^{k}}\pmod*{Q}$ for $0\leq k\leq n-1$ contains $2^{n-1}$ terms of the form $P_{a}^{j}(X)P_{b}^{-j}(X)$. The values of $j$ yielding nonvanishing terms are given in Proposition~\ref{prop:expansionsk}.

\begin{proposition}\label{prop:expansionsk}For\hspace{1.5mm}$0\leq k\leq n-1$, the number of products of the form \mbox{$P_{a}^{j}(X)P_{b}^{-j}(X)$} in the binomial expansion of $\big(P_{a}(X)+P_{b}(X)\big)^{-2^k}$ is $2^{n-1}$. Values of $j$ yielding non-vanishing terms are the integers in $\{0,1,\ldots,2^{n}-1\}$ that are congruent to $0,\ldots, 2^{k}-1$ modulo $2^{k+1}$.
\end{proposition}
\noindent\textbf{Proof.} We have
\begin{align*}
\big(P_{a}(X)+P_{b}(X)\big)^{-2^{k}}&=\sum_{j=0}^{2^{n}-2^{k}-1}{\binom{2^{n}-2^{k}-1}{j}P_{a}^{j}(X)P_{b}^{2^{n}-2^{k}-1-j}(X)}\\
&=P_{b}^{-2^{k}}(X)\sum_{j=0}^{2^{n}-2^{k}-1}{\binom{2^{n}-2^{k}-1}{j}P_{a}^{j}(X)P_{b}^{-j}(X)}.
\end{align*}
Denote by $R_k$ the set containing the values of $0\leq j\leq 2^{n}-2^{k}-1$ such that
\begin{displaymath}
\binom{2^{n}-2^{k}-1}{j}\equiv 1\pmod*{2}.
\end{displaymath}
For any positive integer $t$, let $\nu_2(t)$ be the largest power of $2$ dividing $t$. By examining the parity of binomial coefficients, $R_k$ is the set of $j$'s such that
\begin{displaymath}
\nu_{2}\big((2^{n}-2^{k}-1)!\big)=\nu_{2}\big((2^{n}-2^{k}-1-j)!\big)+\nu_{2}(j!).
\end{displaymath}
We have for $0\leq k\leq n-1$ that
\begin{align}
R_{k}&=\big\{j\in\mathbb{N}\mid 0\leq j\leq 2^{n}-2^{k}-1,\phantom{i}j\equiv 0,1,\ldots,2^{k}-1\pmod*{2^{k+1}}\big\}\nonumber\\
     &=\big\{j\in\mathbb{N}\mid j=2^{k+1}q+r,\phantom{i}q=0,\ldots,2^{n-k-1}-1,\phantom{i}r=0,\ldots,2^{k}-1\big\}\nonumber \label{defnrk}\\
     &=\biguplus_{q=0}^{2^{n-k-1}-1}R_{k,q}\nonumber,
\end{align}
where $R_{k,q}=\big\{j\in\mathbb{N}\mid 0\leq j\leq 2^{n}-1\text{, }j=2^{k+1}q+r\text{, and }r=0,\ldots,2^{k}-1\big\}$. Since $|R_{k,q}|=2^{k}$ and all $R_{k,q}$ are disjoint, then $|R_k|=2^{n-1}$ for all $0\leq k\leq n-1$.\hfill$\blacksquare$

\subsection{Composition identity}\label{app:backforth}

We state an observation that relates permutations as elements of the symmetric group with a bijection over a finite field.
\begin{remark}
From the definition of the permutations $\sigma_{k}$, we observe that
\begin{displaymath}
\big(P_{\sigma_{k}(a)}(X)\big)^{2^{j}}=P_{\sigma_{(k+j)\pmod*{n}}(a)}(X).
\end{displaymath}
\end{remark}

The next proposition concerning the permutations $\sigma_{k}$ provides a cyclically repeated forward-and-backward type composition.

\begin{proposition}\label{backandfor:lemma}
For an even integer $m\geq 2$ and integers $k_j$ such that $0\leq k_j\leq n-1$ for $0\leq j\leq m$, let $\ell$ be defined by
\begin{displaymath}
\ell=\Bigg(\sum_{j=0}^{m}{k_j(-1)^{j\pmod*{2}}}\Bigg)\pmod*{n}.
\end{displaymath}
Then,
\begin{displaymath}
\sigma_{k_m}\sigma_{k_{m-1}}^{-1}\sigma_{k_{m-2}}\cdots\sigma_{k_2}\sigma_{k_1}^{-1}\sigma_{k_0}=\sigma_{\ell}.
\end{displaymath}
\end{proposition}

\noindent\textbf{Proof.} For some irreducible polynomial $Q$, perturbation $P_b$ and by the definition of the $\sigma_{k}$'s, if $0 \leq k \leq n-1$ and $u,v\in \mathbb{F}_{2}^{n}$ are such that $v=\sigma_{k}(u)$, then
\begin{align*}
P_{v}(X)&=P_{\sigma_{k}(u)}(X)=\big(P_{u}(X)+P_{b}(X)\big)^{-2^{k}}\\
&=P^{-2^{n-k}}_{v}(X)+P_{b}(X),\\
P_{u}(X)&=P^{-2^{-k\pmod*{n}}}_{v}(X)+P_{b}(X)\\
&=P_{\sigma_{k}^{-1}(v)}(X).
\end{align*}
Let $c_0, c_1 \in \mathbb{F}_2^n$ be defined by
\begin{align*}
P_{\sigma_{k_0}(a)}(X)&=\big(P_{a}(X)+P_{b}(X)\big)^{-2^{k_0}}= P_{c_0}(X),\\
P_{\sigma_{k_1}^{-1}(c_0)}(X)&=P_{c_0}^{-2^{n-k_1}}(X)+P_{b}(X)\\
&=\Big(\big(P_{a}(X)+P_{b}(X)\big)^{-2^{k_0}}\Big)^{-2^{n-k_1}}+P_{b}(X)\\
&=\big(P_{a}(X)+P_{b}(X)\big)^{2^{n+k_0-k_1}}+P_{b}(X)\\
&=\big(P_{a}(X)+P_{b}(X)\big)^{2^{k_0-k_1}}+P_{b}(X)\\
&= P_{c_1}(X), \text{ and }\\
P_{\sigma_{k_{2}}(c_1)}(X)&=\big(P_{c_1}(X)+P_{b}(X)\big)^{-2^{k_2}}=\Big(\big(P_{a}(X)+P_{b}(X)\big)^{2^{k_0-k_1}}\Big)^{-2^{k_2}}\\
&=\big(P_{a}(X)+P_{b}(X)\big)^{-2^{k_0-k_1+k_2}}.
\end{align*}
Using induction easily completes the proof.\hfill$\blacksquare$

\begin{corollary}
With the same notation as in Proposition~\ref{backandfor:lemma}, for any integer $\ell\geq 0$,
\begin{displaymath}
\sigma_{k_{2i+2}}\big(\sigma_{k_{2i+1}}^{-1}\sigma_{k_{2i}}\cdots\sigma_{k_{1}}^{-1}\sigma_{k_{0}}\big)^{\ell}(a)=\sigma_{(k_{2i+2}+t)\pmod*{n}}(a),
\end{displaymath}
where
\begin{displaymath}
t=\ell\sum_{j=0}^{i}{(k_{2j}-k_{2j+1})}.
\end{displaymath}
\end{corollary}


\section{Properties of strong permutations}\label{sect:specpoly}

{In this section, we analyze properties of the permutations $\sigma$ introduced in Equation~\eqref{eqn:sigma}. In particular, we address the algebraic degree, cycle structure and average number of terms in the algebraic normal form of $\sigma$ that are related to Properties $\mathfrak{P}$1, $\mathfrak{P}$2 and $\mathfrak{P}$3 of Definition~\ref{def:usp}, respectively.}

\begin{remark}\label{rem:perm}
We observe that algebraic normal forms are invariant under permutation of inputs. Hence, to prove Properties $\mathfrak{P}1$ and $\mathfrak{P}3$ for $\sigma = \sigma_{n-1}\cdots \sigma_1\sigma_0$, it is enough to prove them for the final permutation $\sigma_{n-1}$ in the composition.
\end{remark}

In Sections~\ref{sec:P1} and~\ref{sec:P3}, we prove a stronger result than we need given Remark~\ref{rem:perm} and we show the main results in these sections hold
for all $\sigma_k$, $k = 0, \ldots, n-1$.

\subsection{$\mathfrak{P}$1\---Algebraic degree}\label{sec:P1}
For ease of notation, define $\xi_{j,k} = \varphi_{j,2^n-2^k-1}$ for all $j,k$ in
order to match the notation from Equation~\eqref{powermap}. In
this section, we show that $\xi_{j,k}$ have high algebraic degree
for all $j,k$.

\begin{theorem}\label{alg:degthm}
Let
\begin{displaymath}
P_{a}^{2^{n}-2^{k}-1}(X)=\sum_{j=0}^{n-1}{\xi_{j,k}(a)X^{j}}.
\end{displaymath}
Then the algebraic degree of $\xi_{j,k}$ is $n-1$ for all $j,k$.
\end{theorem}

\noindent\textbf{Proof.} We proceed by induction on $k$.
First, let $P_{a}(X)=a_0+\cdots+a_{n-1}X^{n-1}$. For $0\leq k\leq n-1$, and $0\leq j\leq n-1$, we define $\lambda_{j,k}(a)$ as the boolean function which is the coefficient of $X^{j}$ in the expression of $P_{a}^{2^{k}}(X)$. For $k=0$, we have $\lambda_{j,0}(a)=a_{j}$, and for $k=1$, we have
\begin{displaymath}
P_{a}^{2}(X)=\bigg(\sum_{j=0}^{n-1}{a_jX^{j}}\bigg)^{2}=\sum_{j=0}^{n-1}{a_jX^{2j}}=\sum_{j=0}^{n-1}{\lambda_{j,1}(a)X^j}.
\end{displaymath}
The $\lambda_{j,1}$ are linear boolean functions with respect to $a$, since reduction $\pmod*{Q}$ does not increase the degree of the coefficient monomials. Now for $k\geq 1$, if there are linear functions $\lambda_{j,k-1}$ such that \begin{displaymath}
P^{2^{k-1}}_{a}(X)=\sum_{j=0}^{n-1}{\lambda_{j,k-1}(a)X^{j}},
\end{displaymath}
then
\begin{displaymath}
P^{2^k}_{a}(X)=P^{2^{k-1}}_{a}(X)P^{2^{k-1}}_{a}(X)=\sum_{j=0}^{n-1}{\lambda^{2}_{j,k-1}(a)X^{2j}}=\sum_{j=0}^{n-1}{\lambda_{j,k}(a)X^{j}},
\end{displaymath}
where $\lambda_{j,k}$ are linear boolean functions with respect to $a$.

Let $(\ell_0,\ldots,\ell_{n-1})$ be a permutation of $(0, 2, 2^2, \ldots, 2^{n-1})$, and observe that $\sum_{j=0}^{n-1}{\ell_{j}}=2^{n}-2$, then
\begin{align*}
P_{a}^{2^{k}(2^{n}-2)}(X)&=\bigg(\sum_{j=0}^{n-1}{\lambda_{j,k}(a)X^j}\bigg)^{2^{n}-2}=\sum{\binom{2^{n}-2}{\ell_0,\ldots,\ell_{n-1}}\prod_{j=0}^{n-1}{\lambda^{\ell_j}_{j,k}(a)X^{j\ell_{j}}}}\\
&=\sum{\binom{2^{n}-2}{\ell_{0},\ell_{1},\ldots,\ell_{n-1}}\prod_{j=0}^{n-1}{\lambda_{j,k}(a)X^{j\ell_{j}}}}=\sum_{j=0}^{n-1}{\xi_{j,k}(a)X^{j}},
\end{align*}
where without loss of generality we choose $\ell_0=0$, the sum extends over $(\ell_0,\ldots,\ell_{n-1})$ such that $\ell_0+\cdots+\ell_{n-1}=2^{n}-2$,
and $\xi_{j,k}(a)$ have degrees $n-1$ with respect to $a$ for all $j,k$.
Moreover,
\begin{align*}
\nu_{2}\big((2^{n}-2)!\big)&=\sum_{j=1}^{2^{n}-2}{\nu_{2}(j)}=\sum_{j=1}^{2^{n-1}-1}{\nu_{2}(2j)}=\sum_{j=1}^{2^{n-1}-1}{\big(1+\nu_{2}(j)\big)}\\
&=2^{n-1}-1+\sum_{j=1}^{2^{n-1}-1}{\nu_{2}(j)}=2^{n-1}-1+\sum_{j=1}^{2^{n-1}-2}{\nu_{2}(j)}\\
&=2^{n-1}-1+\nu_{2}\big((2^{n-1}-2)!\big)=2^{n-1}-n+\nu_{2}\big((2^{n-1})!\big)\\
&=2^{n}-(n+1)=\sum_{j=1}^{n-1}{\nu_{2}\big((2^{j})!\big)},
\end{align*}
and therefore the multinomial coefficient is odd. Hence the algebraic degree is $n-1$.\hfill$\blacksquare$

\begin{remark} \label{rem:translate}
We observe that since Theorem~\ref{alg:degthm} holds for all
$a\in \mathbb{F}_2^n$, 
the algebraic degrees of $\xi_{j,k}$ corresponding to
$\left(P_a(X) + P_b(X)\right)^{2^n-2^k-1}
= \left(P_{a+b}(X)\right)^{2^n-2^k-1}$ are $n-1$ for all
perturbations $b\in \mathbb{F}_2^n$ of $a$.
\end{remark}

\subsection{$\mathfrak{P}$2\---Period and cycle structure}\label{sec:P2}
In this section, we show that the cycle structure of $\sigma_k$ depends only on the perturbation polynomial, and moreover we show that only $\sigma_0$ can be unicyclic for some perturbation.

\begin{theorem}\label{thm:period}
Let $\ell\geq 0$, $0\leq k\leq n-1$, $1 \leq j \leq \ell$, and $P_b$ be any perturbation. Let
\begin{displaymath}
M_j=\left[
\begin{array}{cc}
0 & 1\\
1 & P_{b}^{ 2^{jk}}
\end{array}\right],
\end{displaymath}
and consider $\ell$ for which
\begin{align}
\prod_{j=1}^{\ell}{M_j}&=\left[\begin{array}{cc}1&0\\0&1\end{array}\right].\label{thmperiod:cond1}
\end{align}
For a large enough positive integer $L$, the period of $\sigma_{k}$ is contained in the sequence of positive integers $\ell\leq L$ satisfying (\ref{thmperiod:cond1}) and
\begin{align}
k\ell &\equiv 0\pmod*{n}\label{thmperiod:cond2}.
\end{align}
\end{theorem}

\noindent\textbf{Proof.} As before, let $Q\in \mathbb{F}_{2}[X]$ be an irreducible polynomial of degree $n$, $P_b$ be a non-zero polynomial, and for $0 \leq k \leq n-1$, let the bijections $\sigma_{k}$ be defined by the rule
\begin{align*}
P_{a}(X)&\mapsto \big(P_{a}(X)+P_{b}(X)\big)^{-2^{k}}\pmod*{Q}.
\end{align*}

For an arbitrary $a\in\mathbb{F}_{2}^{n}$, $\ell\geq 0$, and $0\leq k\leq n-1$, we define
\begin{align*}
P_{\sigma_{k}^{\ell}(a)}(X)&=N_{\ell}(X)D_{\ell}^{-1}(X)\\
&=\big(P_{\sigma_{k}^{\ell-1}(a)}(X)+P_{b}(X)\big)^{-2^{k}},
\end{align*}
where $N_{\ell}\in \mathbb{F}_{2}[X]$, and $D_{\ell}\in \mathbb{F}_{2}[X]$, $D_{\ell}\neq 0$. We note that the case $\ell=0$ corresponds to the identity permutation, and $N_{0}(X)=P_{a}(X)$, and $D_{0}(X)=1$.

Reminiscent to the theory of continued fractions over fields of characteristic $0$, we have for $\ell>0$ that
\begin{align}
\frac{N_{\ell}(X)}{D_{\ell}(X)}&=\bigg(\frac{N_{\ell-1}(X)}{D_{\ell-1}(X)}+P_{b}(X)\bigg)^{-2^{k}}.\label{eqn:partfrac1}
\end{align}

From now on and for readability, we drop the argument $X$ for the elements $N_{\ell}$, $D_{\ell}$ and $P_{b}$. We recall that the period of a permutation $\tau$ is the smallest positive integer $m$ such that $\tau^{m}$ is the identity permutation.

From (\ref{eqn:partfrac1}), we have for $\ell>0$ that
\begin{align*}
N_{\ell}&=D_{\ell-1}^{2^{k}},\\
D_{\ell}&=N_{\ell-1}^{2^{k}}+P_{b}^{2^{k}}D_{\ell-1}^{2^{k}}.
\end{align*}
Equivalently in matrix notation:
\begin{displaymath}
\left[
\begin{array}{c}
N_{\ell}\\
D_{\ell}
\end{array}
\right]=\Bigg(\prod_{j=1}^{\ell}\left[
\begin{array}{cc}
0 & 1\\
1 & P_{b}^{2^{kj}}
\end{array}\right]\Bigg)\left[\begin{array}{c}
N_{0}^{2^{k\ell}}\\
D_{0}^{2^{k\ell}}
\end{array}
\right].
\end{displaymath}

For $j\geq 0$, let $M_j\in\mathbb{F}_{2}[X]^{2\times{}2}$ be
\begin{displaymath}
M_j=\left[
\begin{array}{cc}
0 & 1\\
1 & P_{b}^{ 2^{jk}}
\end{array}\right].
\end{displaymath}
We have $P_{a}^{2^{y}} \equiv P_{a}^{2^{y\pmod*{n}}}$,
$N_{\ell}^{2^{kj}} \equiv N_{\ell}^{2^{kj\pmod*{n}}}$ and
$D_{\ell}^{2^{kj}} \equiv D_{\ell}^{2^{kj\pmod*{n}}}$, for any $a$.
We obtain the cycle structure and the period from the values $\ell\leq L$ such that (\ref{thmperiod:cond1}) and (\ref{thmperiod:cond2}) are satisfied, with $L$ sufficiently large.\hfill$\blacksquare$

\begin{remark}
The matrices $M_{j}$, and thus the product $M_{1}M_{2}\cdots M_{\ell}$ only depend on the perturbation. As a consequence, the period is independent from the input $a$.
\end{remark}

\begin{corollary}\label{cor:giantkeq0}
Only $\sigma_0$ can be unicyclic.
\end{corollary}
\noindent\textbf{Proof.} Recall that $0\leq k<n$. To have a single
cycle of maximal length, Equation (\ref{thmperiod:cond1}) must be satisfied with $\ell=2^{n}$. If $\ell=2^{n}$, then Equation (\ref{thmperiod:cond2}) is only satisfied when $k=0$.\hfill$\blacksquare$

\begin{remark}
Empirically, for a fixed $n$, the proportion of perturbation polynomials leading to unicyclic $\sigma_0$'s is the same for all irreducible polynomials. For instance, we checked experimentally that if $n=2^{2^{\kappa}}$ with $\kappa\in\{1,2,3,4\}$, then this proportion is one half.
\end{remark}

\subsection{$\mathfrak{P}$3\---Average number of terms}\label{sec:P3}
We focus on the third property of a unicyclic strong permutation: the number of terms in the algebraic normal forms of its coordinate functions.
For a positive integer $t$, its \emph{Hamming weight} is the number of ones in its binary expansion. For $0\leq t<2^{n}-1$ with Hamming weight $h$, the expansion of $P_{a}(X)^{t}$ cannot have more than $2^{h+1}-1$ monomials. In our case, $h=n-1$ since we consider powers $2^{n}-2^{k}-1$, and therefore the maximum number of terms is $2^{n}-1$.

In this section we give an asymptotic result that states that the average number of terms in the algebraic normal forms of all $\sigma_k$ is $2^{n-1}$, where the average is taken over all irreducibles $Q(X) \in \mathbb{F}_2[x]$ of degree $n$. In order for the permutation $\sigma = \sigma_{n-1} \cdots \sigma_1 \sigma_0$ from Equation~\eqref{eqn:sigma} to be considered unicyclic strong, all of its component functions must have numbers of terms greater than this average.

We require an assumption on the distribution of the coefficients
of irreducible polynomials over $\mathbb{F}_{2}$ in order to give
our main result of this section.

\paragraph{Assumption 1.} Let $n>0$ and let $1\leq j\leq n-1$.
The number of irreducible polynomials of degree $n$ over
$\mathbb{F}_{2}$ with coefficient of $X^j$ fixed to either
$0$ or to $1$ {tends to} $1/2$ as $n$ tends to $\infty$.

As justification for the assumption, we recall the famous Hansen-Mullen conjecture~\cite{HM}, which was first proven by Wan~\cite[Theorem 1.6]{Wan}.

\begin{theorem}\label{thm:Wan}
Let $q$ be a prime power, let $m$ and $n$ be positive integers with $m \geq n \geq 1$ and let $a \in \mathbb{F}_q$ with $a \neq 0$ if $n = 1$. If either $m \geq 36$ or $q > 19$, then there is a monic irreducible polynomial in $\mathbb{F}_q[X]$ of the form $g(X) = X^m + a_{m-1} X^{m-1} + \cdots + a_n X^n + a_{n-1}X^{n-1} + \cdots + a_1 X + 1$, with $a_{n-1} = a$.
\end{theorem}

Theorem~\ref{thm:Wan} is a result on existence of irreducible polynomials with a given coefficient fixed to any value. Assumption 1 is a reflection of the observation that the result does not depend on the fixed value $a$ for $n > 1$.

We can be more precise than this first-order heuristic. Theorem~\ref{thm:Wan} was proven using character sum techniques. The main technique is to construct a characteristic function for a given quantity, extract the leading term from the trivial character, and apply an estimation (for example, the Weil bound) on the remaining sums to get an expression for the error term. Existence is guaranteed as long as the error term is dominated by the main term.

Wan's proof essentially states that the main term is estimated by $\pi_m/(q-1)$, where $\pi_m$ is the number of irreducible polynomials of degree $m$ over $\mathbb{F}_q$. The details are out of scope of this work. Wan's result applies particularly well when $m$ is large with respect to $n$, and when $q$ is large. If $q$ is large, then $q-1 \approx q$, which supports (a generalization of) Assumption 1. Unfortunately, in our particular case $q=2$, so we further justify Assumption 1 with experiments; see Appendix~\ref{app:nb_zeros_ones}.

\begin{theorem}\label{thm:average}
For $n>0$, and $0\leq k\leq n-1$, let
\begin{displaymath}
P_{a}^{2^n-2^k-1}(X)=\sum_{j=0}^{n-1}{\xi_{j,k}(a)X^{j}}.
\end{displaymath}
Using Assumption $1$, the number of terms in the algebraic normal form of $\xi_{j,k}$ {tends to} $2^{n-1}$.
\end{theorem}

\noindent\textbf{Proof.} We have
\begin{align*}
\big(P_{a}(X)\big)^{2^{n}-2^{k}-1}&=\Bigg(\sum_{i=0}^{n-1}{a_{i}X^{i}}\Bigg)^{2^{n}-2^{k}-1}=\Bigg(\sum_{i=0}^{n-1}{a_{i}X^{i}}\Bigg)^{(2^{n}-2)2^{k}}\\
&=\Bigg(\sum_{i=0}^{n-1}{a_{i}X^{i2^{k}}}\Bigg)^{2^{n}-2}=\Bigg(\sum_{i=0}^{n-1}{\lambda_{i,k}(a)X^{i}}\Bigg)^{2^{n}-2}\\
&=\sum_{\stackrel{(\ell_0,\ldots,\ell_{n-1})}{\ell_0+\cdots+\ell_{n-1}=2^{n}-2}}\binom{2^{n}-2}{\ell_0,\ldots,\ell_{n-1}}\prod_{j=0}^{n-1}{\lambda_{j,k}^{\ell_{j}}(a)X^{j\ell_{j}}}.
\end{align*}
As in the proof of Theorem \ref{alg:degthm}, let $(\ell_0, \ldots,\ell_{n-1})$ be one of the $n!$ permutations of $(0, 2, \ldots, 2^{n-1})$ and observe that $\sum_{j=0}^{n-1}{\ell_{j}}=2^{n}-2$. For convenience, let $\Upsilon$ be the set of vectors of integers for which the multinomial coefficient is odd, and denote by $[P(X)]_{i}\pmod*{Q}$ the coefficient of $X^{i}$ in the expression of $P\pmod*{Q}$. Then,
\begin{align*}
\big(P_{a}(X)\big)^{2^{n}-2^{k}-1}&=\sum_{\ell\in \Upsilon}{\prod_{j=0}^{n-1}{\lambda_{j,k}^{\ell_j}(a)}\sum_{i=0}^{n-1}{[X^{\sum_{j=0}^{n-1}{j\ell_j}}]_{i}X^{i}}}\\
&=\sum_{i=0}^{n-1}{\Bigg(\sum_{\ell\in\Upsilon}{\prod_{j=0}^{n-1}{\lambda_{j,k}^{\ell_j}(a)}[X^{\sum_{j=0}^{n-1}{j\ell_j}}]_{i}}\Bigg)X^{i}}=\sum_{i=0}^{n-1}{\xi_{i,k}(a)X^{i}},
\end{align*}
where
\begin{align*}
\xi_{i,k}(a)&= \sum_{\ell\in\Upsilon}{\prod_{j=0}^{n-1}{\lambda_{j,k}^{\ell_j}(a)}[X^{\sum_{j=0}^{n-1}{j\ell_j}}]_{i}}.
\end{align*}

The quantities $\ell_j$ are distinct powers of two and only one of $\ell_j$ is zero. Hence $\lambda_{j,k}^{\ell_j}(a)=\lambda_{j,k}$ when $\ell_j\neq0$, and $\lambda_{j,k}^{\ell_j}(a)=1$ when $\ell_j= 0$. By Assumption 1, for an arbitrary $\ell\in\Upsilon$, the coefficient $[X^{\sum_{j=0}^{n-1}{j\ell_j}}\pmod*{Q}]_i$ is equally likely to be $0$ or $1$ as $Q$ runs over $\mathcal{I}_{n}$ and $n$ is large. Then the average number of terms in the algebraic normal form of $\xi_{i,k}$ is $2^{n-1}$.\hfill$\blacksquare$

\begin{remark}\label{rem:translate2}
As in Remark~\ref{rem:translate}, we observe that Theorem~\ref{thm:average} holds for all $a \in \mathbb{F}_2^n$, and hence also for all perturbations of $a$.
\end{remark}

\section{Empirical results on strong permutations}\label{sect:compo}

In this section, we present some empirical results on the compositions $\sigma = \sigma_{n-1}\cdots\sigma_{0}$.
In Section~\ref{sec:proportion}, we give some experimental results on the proportions of strong permutations of small degrees. In Sections \ref{sect:numberofsolslineareqn} and \ref{sect:lat}, we give some empirical analysis of the difference table and the linear approximation table for some particular $\sigma$.

\subsection{Permutations satisfying Properties $\mathfrak{P}2$ and $\mathfrak{P}3$}\label{sec:proportion}
By Section~\ref{sec:P1}, the algebraic degree of all $\sigma$ defined as in Equation~\eqref{eqn:sigma} is $n-1$ for all $\sigma$. Since Sections~\ref{sec:P2} and~\ref{sec:P3} do not give guarantees on Properties $\mathfrak{P}2$ and $\mathfrak{P}3$ holding, we present some empirical results on permutations satisfying these properties.

We recall from Section~\ref{sec:permsfromperbs}, for $n=8$, $Q(X)=X^8+X^4+X^3+X+1$ (Rijndael's polynomial) and $t\equiv -1\pmod*{255}$, that exactly half of the possible perturbations lead to a unicyclic permutation. Empirical observations suggest that if $n=2^{2^\kappa}$ for some $\kappa$ and $t\equiv -1\pmod*{2^n-1}$, then there are exactly $2^{n-1}$ unicyclic permutations given by the rule
\begin{displaymath}
P_{a}(X)\mapsto (P_{a}(X)+P_b(X))^{-1}\pmod*{Q}.
\end{displaymath}
When $\kappa=1$, $2$, $3$, and $4$, we empirically verified that the proportion of perturbation polynomials for an arbitrary representation that lead to unicyclic permutations is exactly one half. We stop at $\kappa=4$ given our computational resources, noting that when $\kappa=5$, there are approximately $2^{27}$ irreducible polynomials and
$2^{32}$ perturbation polynomials
yielding approximately 
$2^{59}$ pairs for which the cycle structure must be found.

For $\sigma=\sigma_{n-1}\cdots\sigma_{0}$, 
$n$ even,
and $n \leq 30$, we conducted an exhaustive search over all irreducible polynomials
and all perturbation vectors and found no unicyclic
permutations. We conjecture that this continues to hold.

\begin{cumjacere}\label{conj:evendeg}
If $n$ is even, then the composition $\sigma=\sigma_{n-1}\cdots \sigma_{0}$ is not unicyclic.
\end{cumjacere}

In Table~\ref{tab:conjec_rat}, we run exhaustively through all pairs of irreducible and perturbation polynomials and count for a given perturbation how many irreducible polynomials lead to a unicyclic permutation $\sigma$. For brevity, we report only the minimum and maximum ratios of the number of irreducibles per perturbation.

\begin{table}[h]
\centering
\caption{Min-max ratios of unicyclic permutations of degree $n$.}\label{tab:conjec_rat}
\begin{tabular}{|c|c|}\hline
\multicolumn{2}{|c|}{$n=7$}\\ \hline \hline
Min & $2$ \\ \hline
Max & $14$\\ \hline
$|\mathcal{I}_{7}|$ & $18$\\ \hline
\end{tabular}
\begin{tabular}{|c|c|} \hline
\multicolumn{2}{|c|}{$n=9$}\\ \hline \hline
Min & $2$\\ \hline
Max & $18$\\ \hline
$|\mathcal{I}_{9}|$ & $56$\\ \hline
\end{tabular}
\begin{tabular}{|c|c|} \hline
\multicolumn{2}{|c|}{$n=11$}\\ \hline \hline
Min & $14$\\ \hline
Max & $49$\\ \hline
$|\mathcal{I}_{11}|$ & $186$\\ \hline
\end{tabular}
\end{table}

Let $\mathcal{J}_n$ be the number of irreducible polynomials of degree $n$ such that, for the fixed perturbation polynomial $P_b(X) = X^{n-1} + 1$, the permutation $\sigma$ from Equation~\eqref{eqn:sigma} is unicyclic. Table~\ref{tab:proportions} shows exhaustive results for the ratio $\mathcal{J}_n/\mathcal{I}_n$ for odd $n \leq 25$. We also sampled randomly $4000$ irreducible polynomials of degree $33$ among which $483$ led to unicyclic permutations, and thus resulting in an estimation of $\frac{|\mathcal{J}_{33}|}{|\mathcal{I}_{33}|}\approx 0.12075$.

\begin{longtable}{|c||c|c|c|}
\caption{Ratios $\mathcal{J}_{n}\slash \mathcal{I}_{n}$ when $P_{b}(X)=1+X^{n-1}$.}
\label{tab:proportions}\\
\hline
$n$ & $|\mathcal{J}_{n}|$ & $|\mathcal{I}_{n}|$ & $|\mathcal{J}_{n}|\slash|\mathcal{I}_{n}|$ \\ \hline \hline
\endfirsthead
$n$ & $|\mathcal{J}_{n}|$ & $|\mathcal{I}_{n}|$ & $|\mathcal{J}_{n}|\slash|\mathcal{I}_{n}|$ \\ \hline
\endhead
\hline \multicolumn{4}{c}{Continued on next page}
\endfoot
\endlastfoot
$3$ & $1$ & $2$ & $0.5$ \\
\hline
$5$ & $2$ & $6$ & $0.333333$ \\
\hline

$7$ & $6$ & $18$ & $0.333333$ \\
\hline
$9$ & $10$ & $56$ & $0.178571$ \\
\hline
$11$ & $30$ & $186$ & $0.16129$ \\
\hline
$13$ & $87$ & $630$ & $0.138095$ \\
\hline
$15$ & $259$ & $2182$ & $0.118698$ \\
\hline
$17$ & $1130$ & $7710$ & $0.146563$ \\
\hline
$19$ & $3805$ & $27594$ & $0.137892$ \\
\hline
$21$ & $12551$   & $99858$ &$0.125688$\\
\hline
$23$ & $46290$ & $364722$ & $0.126919$\\
\hline
$25$ & $153976$ & $1342176$ & $0.114721$\\
\hline
\end{longtable}

Based on the experimental evidence in Tables~\ref{tab:conjec_rat} and~\ref{tab:proportions}, we make the following conjecture.
\begin{cumjacere}\label{conj:nonnegfraction}
For odd values of $n$, and $P_{b}$ not constant, 
\begin{displaymath}
\liminf_{n\to\infty}\frac{|\mathcal{J}_{n}|}{|\mathcal{I}_{n}|}\neq 0.
\end{displaymath}
\end{cumjacere}

In order for a permutation $\sigma$ to be ``strong", its component functions must all have at least $2^{n-1}$ terms (that is, it must also satisfy Property $\mathfrak{P}3$). We expect that this property is quite rare, a rough approximation based on Theorem~\ref{thm:average} gives that as $n$ grows, roughly $1/2$ of all component functions will have at least $2^{n-1}$ terms. So, we expect roughly $1/2^n$ unicyclic permutations to be strong.

Table~\ref{tab:USPs} tabulate the results of an exhaustive search for the total number of unicyclic permutations constructed as in Equation~\eqref{eqn:sigma} for $n=7,9,11$ (the search is currently running for $n=13$). 

\begin{longtable}{|c|c|c|}
\caption[supertrong]{Proportions of unicyclic permutations that are ``strong''.}\label{tab:USPs}\\
\hline
$n$ & unicyclic & strong \\ \hline
\endfirsthead
$n$ & unicyclic & strong \\ \hline
\endhead
\hline \multicolumn{3}{c}{Continued on next page}
\endfoot
\endlastfoot
$7$  & $756$   &  $5$  \\ \hline
$9$  & $5040$  &  $3$  \\ \hline
$11$ & $61380$ &  $21$ \\ \hline
$13$ & unknown &  $\geq 10$ \\ \hline
\end{longtable}

\subsection{Examples of difference tables}\label{sect:numberofsolslineareqn}
The \emph{difference table} of a function is a crucial tool in analyzing its resilience against differential cryptanalysis~\cite{Biham1991}. Let $f\colon \mathbb{F}_2^n\to \mathbb{F}_2^n$, and index the rows and columns of a $2^n\times 2^n$ table $\mathcal{D}_f$ with the elements of $\mathbb{F}_2^n$. For $(c,d) \in \mathbb{F}_2^n\times \mathbb{F}_2^n$, the entry at row $c$ and column $d$ of $\mathcal{D}_f$ is the number of pre-images $a \in \mathbb{F}_2^n$ of the expression $f(a+c) + f(a) = d$; that is,
\begin{displaymath}
\mathcal{D}_f(c,d) = \sum_{a\in \mathbb{F}_{2}^{n}}{\mathds{1}\left\{f(a+c)+f(a)-d\right\}},
\end{displaymath}
where $\mathds{1}\{\cdot\}$ denotes the indicator function of $0$. The maximum entry of $\mathcal{D}_f$ is the \emph{differential uniformity} of $f$, and it is desirable for resilience against differential cryptanalysis to have differential uniformity as small as possible.

In fields of characteristic $2$ the differential uniformity of a function is always at least $2$.
Functions with differential uniformity equal to $2$ are \emph{almost perfect nonlinear} (APN). The inverse function $x\mapsto x^{2^n-2}$ is APN when $n$ is odd, and otherwise has differential uniformity $4$; see~\cite[Section 3.1.7]{Carlet2}.

We present summaries of difference tables for $n=7$, $n=9$, and $n=11$. See Appendix \ref{app:leutebokbier} for more statistics. In Table~\ref{tab:Dgiantn7911}, $n$ refers to the degree, $|\mathcal{I}_{n}|$ to the number of irreducible polynomial of degree $n$, and $|\mathcal{T}_{n}|$ to the number of pairs that lead to a unicyclic permutation. We count the number of unicyclic compositions $\sigma_{n-1}\cdots\sigma_{0}$ for possible differentials, and only nonzero counts are reported. The trivial counts for the maximal differential value of $2^n$ are also reported.

\begin{table}[ht]
\centering \begin{footnotesize}
\caption{Differentials over all unicyclic permutations for $n=7, 9, 11$.}\label{tab:Dgiantn7911}
\begin{tabular}{|c|c|}
\hline
\multicolumn{2}{|c|}{$n=7$}\\ \hline
\multicolumn{2}{|c|}{$|\mathcal{I}_{7}|=18$}\\ \hline
\multicolumn{2}{|c|}{$|\mathcal{T}_{7}|=756$}\\ \hline
\hline Differentials & Counts\\ \hline\hline
$0$ & $6545700$\\ \hline
$2$ & $5541102$\\ \hline
$4$ & $292572$\\ \hline
$6$ & $6174$\\ \hline \hline
$128$ & $756$\\ \hline
\end{tabular}
\begin{tabular}{|c|c|}
\hline
\multicolumn{2}{|c|}{$n=9$}\\ \hline
\multicolumn{2}{|c|}{$|\mathcal{I}_{9}|=56$}\\ \hline
\multicolumn{2}{|c|}{$|\mathcal{T}_{9}|=5040$}\\ \hline
Differentials & Counts \\ \hline\hline
$0$ & $673216992$\\ \hline
$2$ & $636750576$\\ \hline
$4$ & $11137392$\\ \hline
$6$ & $95760$\\ \hline \hline
$512$ & $5040$\\ \hline
\end{tabular}
\begin{tabular}{|c|c|}
\hline
\multicolumn{2}{|c|}{$n=11$}\\ \hline
\multicolumn{2}{|c|}{$|\mathcal{I}_{11}|=186$}\\ \hline
\multicolumn{2}{|c|}{$|\mathcal{T}_{11}|=61380$}\\ \hline
\hline Differentials & Counts\\ \hline\hline
$0$ & $129465640194$\\ \hline
$2$ & $127302770628$\\ \hline
$4$ & $676155942$\\ \hline
$6$ & $1751376$\\ \hline \hline
$2048$ & $61380$ \\ \hline
\end{tabular}
\end{footnotesize}
\end{table}

We note that for a given $n$ the sum over all counts divided by $\mathcal{T}_{n}$ is always equal to $2^{n}$.
In our experiments, for $n \leq 13$ the maximum entry in the difference table is at most $6$. Moreover, the differential $6$ occurs rarely. For $n=15$, Example~\ref{ex:hyperplane} from Appendix~\ref{app:leutebokbier} shows a unicyclic permutation that is nearly APN except for a degenerate hyperplane.

%

\subsection{Examples of linear approximation tables}\label{sect:lat}

The \emph{Walsh spectrum} of a function is a crucial tool in analyzing its resilience against linear cryptanalysis~\cite{matsui1993linear}. Let $f\colon \mathbb{F}_2^n \to \mathbb{F}_2^n$ and let $c,d \in \mathbb{F}_2^n\times \mathbb{F}_2^n$. The \emph{Walsh coefficient} $\mathcal{W}_f(c,d)$ of $f$ at $c,d$ is given by
\[ \mathcal{W}_f(c,d) = \sum_{a \in \mathbb{F}_2^n} (-1)^{a\cdot c + d\cdot f(a)},\]
where $\cdot$ denotes the usual scalar product. The values of the Walsh coefficients measure the distance between a given function and the set of affine functions, and hence the table containing the Walsh coefficients is sometimes called the \emph{linear approximation table}. Similarly to the difference table, we highlight the largest entry (in magnitude) in the linear approximation table, and call this the \emph{linearity} of $f$.

We present summaries of linear approximation tables for $n=7$, $n=9$, and $n=11$. See Appendix \ref{app:leutebokbier} for more statistics. In Tables \ref{tab:Lgiantn7}, \ref{tab:Lgiantn9}, and \ref{tab:Lgiantn11}, $n$ refers to the degree, $|\mathcal{I}_{n}|$ to the number of irreducible polynomial of degree $n$, and $|\mathcal{T}_{n}|$ to the number of pairs that lead to a unicyclic permutation. We count the number of unicyclic compositions $\sigma_{n-1}\cdots\sigma_{0}$ for possible correlations, and only nonzero counts are reported. The trivial maximal or minimal values $\pm 2^n$ are also reported as a check.

\begin{longtable}{|c|c|c|c|}
\caption[Lgiantn7]{Distribution of Walsh coefficients over all unicyclic permutations with $n=7$.}\label{tab:Lgiantn7}\\
\hline
\multicolumn{4}{|c|}{$n=7$}\\ \hline
\multicolumn{4}{|c|}{$|\mathcal{I}_{7}|=18$}\\ \hline
\multicolumn{4}{|c|}{$|\mathcal{T}_{7}|=756$}\\ \hline
\hline Coefficients & Counts & Coefficients & Counts\\ \hline\hline
\endfirsthead
\multicolumn{4}{|c|}{$n=7$}\\ \hline
\multicolumn{4}{|c|}{$|\mathcal{I}_{7}|=18$}\\ \hline
\multicolumn{4}{|c|}{$|\mathcal{T}_{7}|=756$}\\ \hline
\hline Coefficients & Counts & Coefficients & Counts\\ \hline\hline
\endhead
\hline \multicolumn{4}{|c|}{Continued on next page}
\endfoot
\endlastfoot
$-32$&$378$ & $4$&$1367226$\\ \hline
$-28$&$17136$ & $8$&$1288224$\\ \hline
$-24$&$140238$ & $12$&$1185786$\\ \hline
$-20$&$486864$ & $16$&$881622$\\ \hline
$-16$&$864360$ & $20$&$478674$\\ \hline
$-12$&$1202796$ & $24$&$139482$\\ \hline
$-8$&$1282176$ & $28$&$15750$\\ \hline
$-4$&$1366344$ & $32$&$378$\\ \hline
$0$&$1668114$ & $128$&$756$\\ \cline{1-4}
\end{longtable}

\newpage
\begin{longtable}{|c|c|c|c|}
\caption[Lgiantn9]{Distribution of Walsh coefficients over all unicyclic permutations with $n=9$.}\label{tab:Lgiantn9}\\
\hline
\multicolumn{4}{|c|}{$n=9$}\\ \hline
\multicolumn{4}{|c|}{$|\mathcal{I}_{9}|=56$}\\ \hline
\multicolumn{4}{|c|}{$|\mathcal{T}_{9}|=5040$}\\ \hline
\hline Coefficients & Counts & Coefficients & Counts\\ \hline\hline
\endfirsthead
\multicolumn{4}{|c|}{$n=9$}\\ \hline
\multicolumn{4}{|c|}{$|\mathcal{I}_{9}|=56$}\\ \hline
\multicolumn{4}{|c|}{$|\mathcal{T}_{9}|=5040$}\\ \hline
\hline Coefficients & Counts & Coefficients & Counts\\ \hline\hline
\endhead
\hline \multicolumn{4}{|c|}{Continued on next page}
\endfoot
\endlastfoot
$-60$&$504$ & $4$&$72997848$\\ \hline
$-56$&$35784$ & $8$&$73843056$\\ \hline
$-52$&$668304$ & $12$&$75178152$\\ \hline
$-48$&$4377744$ & $16$&$72959544$\\ \hline
$-44$&$14541912$ & $20$&$64853208$\\ \hline
$-40$&$29371104$ & $24$&$60175080$\\ \hline
$-36$&$42948864$ & $28$&$57930768$\\ \hline
$-32$&$52182648$ & $32$&$52175592$\\ \hline
$-28$&$58134888$ & $36$&$42539112$\\ \hline
$-24$&$60220944$ & $40$&$29639736$\\ \hline
$-20$&$64415736$ & $44$&$14436576$\\ \hline
$-16$&$72875880$ & $48$&$4386816$\\ \hline
$-12$&$75241656$ & $52$&$689976$\\ \hline
$-8$&$73638432$ & $56$&$50904$\\ \hline
$-4$&$73057824$ & $512$&$5040$\\ \hline
$0$&$77632128$ & \multicolumn{2}{|c}{}\\ \cline{1-2}
\end{longtable}

\begin{longtable}{|c|c|c|c|}
\caption[Lgiantn11]{Distribution of Walsh coefficients over all unicyclic permutations with $n=11$.}\label{tab:Lgiantn11}\\
\hline
\multicolumn{4}{|c|}{$n=11$}\\ \hline
\multicolumn{4}{|c|}{$|\mathcal{I}_{11}|=186$}\\ \hline
\multicolumn{4}{|c|}{$|\mathcal{T}_{11}|=61380$}\\ \hline
\hline Coefficients & Counts & Coefficients & Counts\\ \hline\hline
\endfirsthead
\multicolumn{4}{|c|}{$n=11$}\\ \hline
\multicolumn{4}{|c|}{$|\mathcal{I}_{11}|=56$}\\ \hline
\multicolumn{4}{|c|}{$|\mathcal{T}_{11}|=61380$}\\ \hline
\hline Coefficients & Counts & Coefficients & Counts\\ \hline\hline
\endhead
\hline \multicolumn{4}{|c|}{Continued on next page}
\endfoot
\endlastfoot
$-108$&$10230$ & $4$&$7190926842$\\ \hline
$-104$&$777480$ & $8$&$7251819894$\\ \hline
$-100$&$17145480$ & $12$&$7065347454$\\ \hline
$-96$&$141090114$ & $16$&$7134428598$\\ \hline
$-92$&$572151624$ & $20$&$7557831930$\\ \hline
$-88$&$1386917928$ & $24$&$7312983018$\\ \hline
$-84$&$2406924630$ & $28$&$6648708198$\\ \hline
$-80$&$3331962150$ & $32$&$6391800162$\\ \hline
$-76$&$3885243516$ & $36$&$6449828814$\\ \hline
$-72$&$4243987110$ & $40$&$6482409318$\\ \hline
$-68$&$4616684424$ & $44$&$6421456932$\\ \hline
$-64$&$4985688708$ & $48$&$6359592030$\\ \hline
$-60$&$5358269400$ & $52$&$6104515164$\\ \hline
$-56$&$5710862718$ & $56$&$5702230644$\\ \hline
$-52$&$6099580212$ & $60$&$5362862670$\\ \hline
$-48$&$6361061058$ & $64$&$4989750018$\\ \hline
$-44$&$6425397528$ & $68$&$4611074292$\\ \hline
$-40$&$6479716782$ & $72$&$4247581932$\\ \hline
$-36$&$6449734698$ & $76$&$3887400000$\\ \hline
$-32$&$6387153696$ & $76$&$3887400000$\\ \hline
$-28$&$6649688232$ & $80$&$3331630698$\\ \hline
$-24$&$7306112550$ & $84$&$2405177346$\\ \hline
$-20$&$7567429716$ & $88$&$1384610040$\\ \hline
$-16$&$7135218354$ & $92$&$571769022$\\ \hline
$-12$&$7062515790$ & $96$&$140161230$\\ \hline
$-8$&$7246246590$ & $100$&$17010444$\\ \hline
$-4$&$7188592356$ & $104$&$789756$\\ \hline
$0$&$7406442252$ & $108$&$16368$\\ \hline
\multicolumn{2}{c|}{} & $2048$&$61380$\\\cline{3-4}
\end{longtable}

We observe that the Walsh coefficients of our unicyclic permutations are tightly compacted around $0$, especially in comparison to a uniform random permutation as in Appendix \ref{app:leutebokbier}. As expected, the weighted average of the Walsh coefficients is $0$.

\section{Conclusion}\label{sect:conclusion}
In this paper, we give permutations over $\mathbb{F}_{2}^{n}$ that \emph{simultaneously} satisfy at least three properties of interest in cryptography: high algebraic degree and large number of terms in the algebraic normal forms of their component functions, and maximal cycle size. Experimentally,  we also find that these permutations have properties that may indicate good resistance against linear and differential cryptanalysis.

It would be interesting to study the distributions of the coefficients of the difference table and linear approximation table of strong unicyclic permutations. Other future work includes determining conditions under which our composition from Section \ref{sect:compo} is unicyclic; this would imply a proof of Conjecture \ref{conj:nonnegfraction} and would likely explain the dichotomy in the cycle structures between odd and even degrees. Another path for future work is to obtain analytic conditions for when a unicyclic permutation has large number of terms in the algebraic normal forms of its component functions.

\newcommand{\SortNoop}[1]{}

\section*{Acknowledgement}

The authors are grateful for the very careful reviews and the constructive suggestions received from the referees.

\appendix
\section{Empirical evidence for Assumption 1}\label{app:nb_zeros_ones}

In Tables \ref{table1:assumptionuno}, \ref{table2:assumptionuno}, \ref{table3:assumptionuno}, and \ref{table4:assumptionuno}, the column entitled $i$ refers to the coefficients of $X^{i}$ of an irreducible polynomial of degree $d$. The column entitled ``Ratios'' gives the ratios of the number of irreducible polynomials of degree $d$ with a $X^{i}$ term.

\begin{longtable}{|c|c||c|c||c|c||c|c|}
\caption[table1:for_assumption_one]{Average proportions by terms of irreducible polynomials.}\label{table1:assumptionuno}\\
\hline
\multicolumn{2}{|c||}{$d = 2$} & \multicolumn{2}{|c||}{$d = 3$} & \multicolumn{2}{|c||}{$d = 4$} & \multicolumn{2}{|c|}{$d = 5$} \\ \hline
$i$ & Ratios & $i$ & Ratios & $i$ & Ratios & $i$ & Ratios \\ \hline \hline
\endfirsthead
\multicolumn{2}{|c||}{$d = 2$} & \multicolumn{2}{|c||}{$d = 3$} & \multicolumn{2}{|c||}{$d = 4$} & \multicolumn{2}{|c|}{$d = 5$} \\ \hline
$i$ & Ratios & $i$ & Ratios & $i$ & Ratios & $i$ & Ratios \\ \hline \hline
\endhead
\hline \multicolumn{4}{|c|}{Continued on next page}
\endfoot
\endlastfoot
$1$ & $1.000000$              & $1$ & $0.500000$ & $1$ & $0.666667$       & $1$ & $0.500000$ \\ \cline{1-8}
\multicolumn{2}{c|}{}         & $2$ & $0.500000$ & $2$ & $0.333333$       & $2$ & $0.666667$ \\ \cline{3-8}
\multicolumn{4}{c|}{}         &                    $3$ & $0.666667$       & $3$ & $0.666667$ \\ \cline{5-8}
\multicolumn{6}{c|}{}         &                                             $4$ & $0.500000$ \\ \cline{7-8}
\end{longtable}

\begin{longtable}{|c|c||c|c||c|c||c|c|}
\caption[table2:for_assumption_one]{Average proportions by terms of irreducible polynomials (continued).}\label{table2:assumptionuno}\\
\hline
\multicolumn{2}{|c||}{$d = 6$} & \multicolumn{2}{|c||}{$d = 7$} & \multicolumn{2}{|c||}{$d = 8$} & \multicolumn{2}{|c|}{$d = 9$} \\ \hline
$i$ & Ratios & $i$ & Ratios & $i$ & Ratios & $i$ & Ratios \\ \hline \hline
\endfirsthead
\multicolumn{2}{|c||}{$d = 6$} & \multicolumn{2}{|c||}{$d = 7$} & \multicolumn{2}{|c||}{$d = 8$} & \multicolumn{2}{|c|}{$d = 9$} \\ \hline
$i$ & Ratios & $i$ & Ratios & $i$ & Ratios & $i$ & Ratios \\ \hline \hline
\endhead
\hline \multicolumn{8}{|c|}{Continued on next page}
\endfoot
\endlastfoot
$1$ & $0.555556$ & $1 $ & $0.500000$ & $1$ & $0.533333$ & $1 $ & $0.500000$\\ \cline{1-8}
$2$ & $0.444444$ & $2 $ & $0.500000$ & $2$ & $0.533333$ & $2 $ & $0.464286$\\ \cline{1-8}
$3$ & $0.333333$ & $3 $ & $0.500000$ & $3$ & $0.600000$ & $3 $ & $0.517857$\\ \cline{1-8}
$4 $ & $0.444444$ & $4 $ & $0.500000$ & $4 $ & $0.533333$ & $4 $ & $0.517857$\\ \cline{1-8}
$5$ & $0.555556$ & $ 5 $ &  $0.500000$ & $ 5 $ & $0.600000$ & $ 5 $ & $0.517857$\\ \cline{1-8}
\multicolumn{2}{c|}{} & $ 6 $ & $0.500000$ & $6 $ & $0.533333$ & $ 6 $ & $0.517857$\\ \cline{3-8}
\multicolumn{4}{c|}{} & $ 7 $ & $0.533333$ & $ 7 $ & $0.464286$\\ \cline{5-8}
\multicolumn{6}{c|}{} & $ 8 $ & $0.500000$\\ \cline{7-8}
\end{longtable}

\begin{longtable}{|c|c||c|c||c|c||c|c|}
\caption[table3:for_assumption_one]{Average proportions by terms of irreducible polynomials (continued).}\label{table3:assumptionuno}\\
\hline
\multicolumn{2}{|c||}{$d = 10$} & \multicolumn{2}{|c||}{$d = 11$} & \multicolumn{2}{|c||}{$d = 12$} & \multicolumn{2}{|c|}{$d = 13$} \\ \hline
$i$ & Ratios & $i$ & Ratios & $i$ & Ratios & $i$ & Ratios \\ \hline \hline
\endfirsthead
\multicolumn{2}{|c||}{$d = 10$} & \multicolumn{2}{|c||}{$d = 11$} & \multicolumn{2}{|c||}{$d = 12$} & \multicolumn{2}{|c|}{$d = 13$} \\ \hline
$i$ & Ratios & $i$ & Ratios & $i$ & Ratios & $i$ & Ratios \\ \hline \hline
\endhead
\hline \multicolumn{8}{|c|}{Continued on next page}
\endfoot
\endlastfoot
$1 $ & $0.515152$ & $1 $ & $0.500000$ & $1 $ & $0.507463$ & $ 1 $ & $0.500000$\\ \hline
$2 $ & $0.515152$ & $2 $ & $0.500000$ & $2 $ & $0.492537$ & $ 2 $ & $0.507937$\\ \hline
$3 $ & $0.484848$ & $3 $ & $0.500000$ & $3 $ & $0.507463$ & $3 $ & $0.507937$\\ \hline
$4 $ & $0.474747$ & $4 $ & $0.516129$ & $4 $ & $0.498507$ & $4 $ & $0.500000$\\ \hline
$5 $ & $0.494949$ & $5 $ & $0.516129$ & $5 $ & $0.477612$ & $5 $ & $0.512698$\\ \hline
$6 $ & $0.474747$ & $6 $ & $0.516129$ & $6 $ & $0.504478$ & $6 $ & $0.495238$\\ \hline
$7 $ & $0.484848$ & $7 $ & $0.516129$ & $7 $ & $0.477612$ & $7 $ & $0.495238$\\ \hline
$8 $ & $0.515152$ & $8 $ & $0.500000$ & $8 $ & $0.498507$ & $8 $ & $0.512698$\\ \hline
$9 $ & $0.515152$ & $9 $ & $0.500000$ & $9 $ & $0.507463$ & $9 $ & $0.500000$\\ \hline
\multicolumn{2}{c|}{} &  $10$ & $0.500000$ & $10 $ & $0.492537$ & $10 $ & $0.507937$\\ \cline{3-8}
\multicolumn{4}{c|}{} & $11 $ & $0.507463$ & $11 $ & $0.507937$ \\ \cline{5-8}
\multicolumn{6}{c|}{} & $12$ & $0.500000$\\ \cline{7-8}
\end{longtable}
In order to make this article concise, we jump from degree $13$ to $26$ up to $29$ inclusively.
\begin{longtable}{|c|c||c|c||c|c||c|c|}
\caption[table4:for_assumption_one]{Average proportions by terms of irreducible polynomials (continued).}\label{table4:assumptionuno}\\
\hline
\multicolumn{2}{|c||}{$d = 26$} & \multicolumn{2}{|c||}{$d = 27$} & \multicolumn{2}{|c||}{$d = 28$} & \multicolumn{2}{|c|}{$d = 29$} \\ \hline
$i$ & Ratios & $i$ & Ratios & $i$ & Ratios & $i$ & Ratios \\ \hline \hline
\endfirsthead
\multicolumn{2}{|c||}{$d = 26$} & \multicolumn{2}{|c||}{$d = 27$} & \multicolumn{2}{|c||}{$d = 28$} & \multicolumn{2}{|c|}{$d = 29$} \\ \hline
$i$ & Ratios & $i$ & Ratios & $i$ & Ratios & $i$ & Ratios \\ \hline \hline
\endhead
\hline \multicolumn{8}{|c|}{Continued on next page}
\endfoot
\endlastfoot
$ 1 $ & $0.500061$ & $1 $ & $0.500000$ & $ 1 $ & $0.500031$ & $ 1 $ & $0.500000$ \\ \cline{1-8}
$ 2 $ & $0.500061$ & $ 2 $ & $0.500000$ & $  2 $ & $0.499969$ & $ 2 $ & $0.500031$ \\ \cline{1-8}
$ 3 $ & $0.499939$ & $ 3 $ & $0.500092$ & $3 $ & $0.500031$ & $3 $ & $0.500031$ \\ \cline{1-8}
$ 4 $ & $0.500036$ & $ 4 $ & $0.500118$ & $4 $ & $0.500063$ & $4 $ & $0.499945$ \\ \cline{1-8}
$ 5 $ & $0.500098$ & $ 5 $ & $0.499878$ & $5 $ & $0.499957$ & $5 $ & $0.500007$ \\ \cline{1-8}
$ 6 $ & $0.499999$ & $ 6 $ & $0.500101$ & $6 $ & $0.499967$ & $6 $ & $0.499954$ \\ \cline{1-8}
$ 7 $ & $0.500038$ & $ 7 $ & $0.500101$ & $7 $ & $0.500003$ & $7 $ & $0.499999$ \\ \cline{1-8}
$ 8 $ & $0.499982$ & $  8 $ & $0.500219$ & $ 8 $ & $0.500044$ & $8 $ & $0.499960$ \\ \cline{1-8}
$ 9 $ & $0.500092$ & $ 9 $ & $0.499944$ & $ 9 $ & $0.499936$ & $9 $ & $0.500074$ \\ \cline{1-8}
$ 10 $ & $0.499886$ & $ 10 $ & $0.500054$ & $ 10 $ & $0.500089$ & $10 $ & $0.500139$ \\ \cline{1-8}
$ 11 $ & $0.500013$ & $ 11 $ & $0.499764$ & $ 11 $ & $0.499940$ & $11 $ & $0.500091$ \\ \cline{1-8}
$ 12 $ & $0.499996$ & $ 12 $ & $0.500161$ & $ 12 $ & $0.499999$ & $12 $ & $0.500012$ \\ \cline{1-8}
$ 13 $ & $0.499994$ & $  13 $ & $0.500063$ & $ 13 $ & $0.499984$ & $13 $ & $0.499931$ \\ \cline{1-8}
$ 14 $ & $0.499996$ & $ 14 $ & $0.500063$ & $ 14 $ & $0.499982$ & $14 $ & $0.499990$ \\ \cline{1-8}
$ 15 $ & $0.500013$ & $ 15 $ & $0.500161$ & $  15 $ & $0.499984$ & $15 $ & $0.499990$ \\ \cline{1-8}
$ 16 $ & $0.499886$ & $ 16 $ & $0.499764$ & $ 16 $ & $0.499999$ & $ 16 $ & $0.499931$ \\ \cline{1-8}
$ 17 $ & $0.500092$ & $ 17 $ & $0.500054$ & $ 17 $ & $0.499940$ & $17 $ & $0.500012$ \\ \cline{1-8}
$ 18 $ & $0.499982$ & $ 18 $ & $0.499944$ & $ 18 $ & $0.500089$ & $ 18 $ & $0.500091$ \\ \cline{1-8}
$ 19 $ & $0.500038$ & $ 19 $ & $0.500219$ & $ 19 $ & $0.499936$ & $ 19 $ & $0.500139$ \\ \cline{1-8}
$ 20 $ & $0.499999$ & $ 20 $ & $0.500101$ & $ 20 $ & $0.500044$ & $ 20 $ & $0.500074$ \\ \cline{1-8}
$ 21 $ & $0.500098$ & $ 21 $ & $0.500101$ & $ 21 $ & $0.500003$ & $ 21 $ & $0.499960$ \\ \cline{1-8}
$ 22 $ & $0.500036$ & $ 22 $ & $0.499878$ & $22 $ & $0.499967$ & $ 22 $ & $0.499999$ \\ \cline{1-8}
$ 23 $ & $0.499939$ & $ 23 $ & $0.500118$ & $23 $ & $0.499957$ & $ 23 $ & $0.499954$ \\ \cline{1-8}
$ 24 $ & $0.500061$ & $ 24 $ & $0.500092$ & $24 $ & $0.500063$ & $ 24 $ & $0.500007$ \\ \cline{1-8}
$ 25 $ & $0.500061$ & $ 25 $ & $0.50000$ & $25 $ & $0.500031$ & $ 25 $ & $0.499945$ \\ \cline{1-8}
\multicolumn{2}{c|}{} &  $26 $ & $0.500000$ & $ 26 $ & $0.499969$ & $ 26 $ & $0.500031$ \\ \cline{3-8}
\multicolumn{4}{c|}{} &  $27 $ & $0.500031$ & $ 27 $ & $0.500031$\\ \cline{5-8}
\multicolumn{6}{c|}{} &  $28 $ & $0.500000$\\ \cline{7-8}
\end{longtable}

\section{Differentials and correlations}\label{app:leutebokbier}

Example~\ref{ex:hyperplane} gives two unicyclic permutations
that are APN except for a fixed set of $2^{15}$ pairs giving differentials of size $2^{15}$.

\begin{example}\label{ex:hyperplane}
The following two examples give unicyclic permutations $\sigma$ that are APN aside from $2^{15}$ $(c,d)$ pairs each giving $2^{15}$ solutions to $\sigma(X\oplus c) \oplus \sigma(X) = d$. A summary of both difference tables is given in Table~\ref{tab:degenerate}. We present these examples for interest and further study, but we make no claims on their suitability for cryptography. 
\begin{longtable}{|c|c|}\caption{Summary of the difference tables of two unicyclic strong permutations from Equations~\eqref{eqn:Q1} and~\eqref{eqn:Q2}}\label{tab:degenerate}
\\
\hline
Differentials & Counts \\ \hline\hline
\endfirsthead
\hline Differentials & Counts\\ \hline\hline
\endhead
\hline \multicolumn{2}{|c|}{Continued on next page}
\endfoot
\endlastfoot
$0$        &   $1073250409$  \\ \hline
$2$        &   $458647$      \\ \hline
$32768$    &   $32768 $      \\ \hline
\end{longtable}

\begin{align}
Q_1(X)     & = 
              1 + X + X^7 + X^{10} + X^{15} \label{eqn:Q1} \\
P_{1,b}(X) & = 
              1  + X^3 + X^5 + X^7 + X^{11} + X^{12} + X^{13} \nonumber \\
Q_2(X)     &= 
             1 + X^2 + X^3 + X^7 + X^8 + X^{12} + X^{13} + X^{14} + X^{15} \label{eqn:Q2}\\
P_{2,b}(X) &= 
                 X^2 + X^5 + X^6 + X^7 + X^8 + X^{11} + X^{12} + X^{14} \nonumber
\end{align}
\end{example}

\begin{example}\label{diftab1}
Let $n=17$ and let $Q$ and $P_b$ be as follows,
\begin{align*}
Q(X) &= 1 + X + X^4 + X^8 + X^{11} + X^{12} + X^{13} + X^{14} + X^{15} + X^{16} + X^{17}, \\
P_b(X) &= 1 + X^{16}.
\end{align*}
The difference table for $\sigma$ defined with $Q$ and $P_b$ is
\begin{longtable}{|c|c|}
\hline
Differentials & Counts \\ \hline\hline
\endfirsthead
\hline Differentials & Counts\\ \hline\hline
\endhead
\hline \multicolumn{2}{|c|}{Continued on next page}
\endfoot
\endlastfoot
$0$ & $8591113477$\\
\hline
$2$ & $8587642420$\\
\hline
$4$ & $1113222$\\
\hline
$6$ & $64$\\
\hline
$131072$ & $1$\\
\hline
\end{longtable}
\end{example}

\begin{example}\label{diftab2}
Let $n=19$ and let $Q$ and $P_b$ be as follows,
\begin{align*}
Q(X) &= 1 + X^5 + X^7 + X^8 + X^9 + X^{11} + X^{13} + X^{16} + X^{17} + X^{18} + X^{19},\\
P_b(X) &= 1 + X^{18}.
\end{align*}
The difference table for $\sigma$ defined with $Q$ and $P_b$ is
\begin{longtable}{|c|c|}
\hline
Differentials & Counts \\ \hline\hline
\endfirsthead
\hline Differentials & Counts\\ \hline\hline
\endhead
\hline \multicolumn{2}{|c|}{Continued on next page}
\endfoot
\endlastfoot
$0$ & $137444193323$\\
\hline
$2$ & $137428735987$\\
\hline
$4$ & $4977558$\\
\hline
$6$ & $75$\\
\hline
$524288$ & $1$\\
\hline
\end{longtable}
\end{example}

Example~\ref{example:lat} gives a comparison of a particular unicyclic permutation with a uniformly randomly sampled permutation.

\begin{example}\label{example:lat}
Let $n=15$ and let $Q$ and $P_b$ be as follows,
\begin{align*}
Q(X) &= 1 + X^3 + X^4 + X^5 + X^7 + X^{14} + X^{15}, \\
P_b(X) &= 1 + X^{14}.
\end{align*}
%
The linear approximation table for $\sigma$ defined with $Q$ and $P_b$ is
\begin{longtable}{|c|c||c|c|}
\hline
Coefficients  & Counts & Coefficients  & Counts\\ \hline \hline
\endfirsthead
Coefficients  & Counts & Coefficients  & Counts\\ \hline
\endhead
\hline \multicolumn{4}{|c|}{Continued on next page}
\endfoot
\endlastfoot
$-384$ & $6$ (extreme) & $4$ & $7419616$\\ \hline
$-380$ & $146$ & $8$ & $7521798$\\ \hline
\vdots & \vdots & $12$ & $7751075$\\ \hline
$-12$ & $7748469$ & \vdots & \vdots\\ \hline
$-8$ & $7519934$ & $380$ & $148$\\ \hline
$-4$ & $7416332$ & $384$ & $4$ (extreme)\\ \hline
$0$ & $7486434$ & $32768$ & $1$ (trivial)\\ \hline
               \end{longtable}

Comparing with a uniform random permutation, we obtain

\begin{longtable}{|c|c||c|c|}
\hline
Coefficients  & Counts & Coefficients  & Counts \\ \hline \hline
\endfirsthead
Coefficients  & Counts & Coefficients  & Counts \\ \hline
\endhead
\hline \multicolumn{4}{|c|}{Continued on next page}
\endfoot
\endlastfoot
$-1088$ & $1$ (extreme) & $4$ & $9464656$\\ \hline
$-1072$ & $1$ & \vdots & \vdots\\ \hline
\vdots & \vdots & $380$ & $1046219$\\ \hline
$-384$ & $998280$ & $384$ & $997790$ \\ \hline
$-380$ & $1044401$ & \vdots & \vdots\\ \hline
\vdots & \vdots & $1196$ & $1$ \\ \hline
$-4$ & $9465140$ & $1252$ & $1$ (extreme)\\ \hline
$0$ & $9525299$ & $32768$ & $1$ (trivial)\\ \hline
\end{longtable}
\end{example}

\section{An example with intermediate round computations}\label{sect:intermediate_comp}

For $1\leq i\leq n$, let
\begin{align*}
P_{a^{(i)}}(X)&=\big(P_{a^{(i-1)}}(X)+P_{b}(X)\big)^{-2^{i-1}}.
\end{align*}
In Table \ref{bouletteD6}, we give the sequence $a=a^{(0)}\to a^{(1)}\to \cdots \to a^{(n)}$; columns entitled $a^{(i)}$ contain the output of the partial computations $a^{(i)}=\sigma_{i-1}\sigma_{i-2}\cdots\sigma_{0}$. In each column, we give an underlined boldfaced entry that signifies a cycle that is not of maximal length up to the given column/round. For instance, after two rounds, we have a fixed point since $\sigma_{1}\sigma_{0}(59)=59$, and after four rounds we have a cycle of length three since $\sigma_{3}\sigma_{2}\sigma_{1}\sigma_{0}(0)=23$, $\sigma_{3}\sigma_{2}\sigma_{1}\sigma_{0}(23)=35$, and $\sigma_{3}\sigma_{2}\sigma_{1}\sigma_{0}(35)=0$.

\newpage

\begin{center}
\begin{longtable}{p{4mm}||p{4mm}|p{4mm}|p{4mm}|p{4mm}|p{4mm}|p{4mm}||||p{4mm}||p{4mm}|p{4mm}|p{4mm}|p{4mm}|p{4mm}|p{4mm}}
\caption[A degree $6$ example]{Intermediate round computations for $P_{b}(X)=1+X^{5}$ and $Q(X)=1+X+X^4+X^5+X^6$.} \label{bouletteD6} \\

$a$ & $a^{(1)}$ & $a^{(2)}$ & $a^{(3)}$ & $a^{(4)}$ & $a^{(5)}$ & $a^{(6)}$ & $a$ & $a^{(1)}$ & $a^{(2)}$ & $a^{(3)}$ & $a^{(4)}$ & $a^{(5)}$ & $a^{(6)}$ \\ \hline \hline
\endfirsthead

$a$ & $a^{(1)}$ & $a^{(2)}$ & $a^{(3)}$ & $a^{(4)}$ & $a^{(5)}$ & $a^{(6)}$ & $a$ & $a^{(1)}$ & $a^{(2)}$ & $a^{(3)}$ & $a^{(4)}$ & $a^{(5)}$ & $a^{(6)}$\\ \hline
\endhead
\endfoot

\endlastfoot

$ 0$ & $10$ & $38$ & $39$ & $\underline{\textbf{23}}$ & $19$ & $39$ & $32$ & $ 1$ & $36$ & $24$ & $37$ & $\underline{\textbf{18}}$ & $35$ \\
\hline
$ 1$ & $13$ & $ 5$ & $29$ & $53$ & $34$ & $48$ & $33$ & $ 0$ & $49$ & $32$ & $ 1$ & $16$ & $\underline{\textbf{33}}$ \\
\hline
$ 2$ & $46$ & $13$ & $17$ & $17$ & $63$ & $30$ & $34$ & $44$ & $18$ & $34$ & $24$ & $ 3$ & $41$\\
\hline
$ 3$ & $38$ & $40$ & $53$ & $43$ & $27$ & $36$ & $35$ & $58$ & $25$ & $33$ & $\underline{\textbf{ 0}}$ & $56$ & $27$\\
\hline
$ 4$ & $48$ & $24$ & $12$ & $14$ & $50$ & $10$ & $36$ & $55$ & $20$ & $52$ & $21$ & $53$ & $22$\\
\hline
$ 5$ & $18$ & $43$ & $59$ & $36$ & $48$ & $55$ & $37$ & $29$ & $61$ & $25$ & $19$ & $40$ & $53$\\
\hline
$ 6$ & $47$ & $10$ & $ \underline{\textbf{6}}$ & $39$ & $43$ & $14$ & $38$ & $50$ & $45$ & $44$ & $29$ & $61$ & $45$\\
\hline
$ 7$ & $34$ & $55$ & $46$ & $ 2$ & $ 6$ & $21$ &  $39$ & $22$ & $17$ & $ 5$ & $26$ & $10$ & $51$\\
\hline
$ 8$ & $43$ & $19$ & $47$ & $45$ & $57$ & $17$ &$40$ & $61$ & $56$ & $19$ & $50$ & $25$ & $28$\\
\hline
$ 9$ & $21$ & $53$ & $23$ & $33$ & $ 0$ & $25$ & $41$ & $52$ & $39$ & $35$ & $15$ & $22$ & $ 3$\\
\hline
$10$ & $41$ & $60$ & $62$ & $31$ & $29$ & $ 8$ &  $42$ & $12$ & $28$ & $60$ & $13$ & $12$ & $19$ \\
\hline
$11$ & $20$ & $ 9$ & $50$ & $56$ & $28$ & $60$ & $43$ & $33$ & $ 0$ & $45$ & $55$ & $41$ & $ 9$\\
\hline
$12$ & $59$ & $62$ & $30$ & $44$ & $26$ & $49$ & $44$ & $32$ & $ 1$ & $ 2$ & $38$ & $21$ & $61$ \\
\hline
$13$ & $ 3$ & $ 6$ & $40$ & $61$ & $\underline{\textbf{49}}$ & $29$ & $45$ & $11$ & $46$ & $36$ & $11$ & $38$ & $47$ \\
\hline
$14$ & $39$ & $42$ & $ 3$ & $46$ & $ 4$ & $11$ & $46$ & $62$ & $31$ & $18$ & $22$ & $37$ & $58$\\
\hline
$15$ & $35$ & $29$ & $ 8$ & $42$ & $17$ & $12$ & $47$ & $25$ & $14$ & $21$ & $ 8$ & $35$ & $18$\\
\hline
$16$ & $19$ & $21$ & $61$ & $10$ & $46$ & $\underline{\textbf{16}}$ & $48$ & $57$ & $12$ & $27$ & $62$ & $30$ & $57$ \\
\hline
$17$ & $37$ & $26$ & $56$ & $59$ & $ 2$ & $20$ & $49$ & $26$ & $54$ & $38$ & $ 7$ & $\underline{\textbf{23}}$ & $59$\\
\hline
$18$ & $23$ & $27$ & $16$ & $27$ & $\underline{\textbf{13}}$ & $37$ &  $50$ & $49$ & $15$ & $43$ & $28$ & $52$ & $50$\\
\hline
$19$ & $ 7$ & $22$ & $63$ & $30$ & $55$ & $38$ & $51$ & $36$ & $57$ & $37$ & $32$ & $ 1$ & $62$ \\
\hline
$20$ & $60$ & $16$ & $28$ & $ 9$ & $39$ & $34$ & $52$ & $40$ & $ 8$ & $22$ & $12$ & $33$ & $ 0$\\
\hline
$21$ & $ 8$ & $34$ & $57$ & $ 3$ & $51$ & $13$ &  $53$ & $42$ & $37$ & $15$ & $34$ & $11$ & $ 6$ \\
\hline
$22$ & $ 5$ & $58$ & $10$ & $20$ & $ 9$ & $ 7$ & $54$ & $51$ & $ 2$ & $41$ & $52$ & $47$ & $56$\\
\hline
$23$ & $28$ & $52$ & $ 7$ & $\underline{\textbf{35}}$ & $\underline{\textbf{32}}$ & $ 1$ & $55$ & $ 6$ & $50$ & $54$ & $ 6$ & $ 7$ & $43$\\
\hline
$24$ & $17$ & $ 3$ & $20$ & $60$ & $36$ & $44$ & $56$ & $14$ & $ 7$ & $42$ & $ 5$ & $15$ & $42$\\
\hline
$25$ & $27$ & $ 4$ & $55$ & $51$ & $62$ & $31$ & $57$ & $63$ & $30$ & $48$ & $48$ & $44$ & $15$\\
\hline
$26$ & $45$ & $48$ & $11$ & $41$ & $60$ & $ 2$ & $58$ & $56$ & $33$ & $ 0$ & $54$ & $20$ & $52$\\
\hline
$27$ & $ 2$ & $51$ & $ 4$ & $57$ & $ 5$ & $32$ & $59$ & $16$ & $\underline{\textbf{59}}$ & $13$ & $63$ & $31$ & $26$\\
\hline
$28$ & $ 9$ & $47$ & $49$ & $18$ & $42$ & $63$ & $60$ & $ 4$ & $44$ & $58$ & $49$ & $58$ & $54$ \\
\hline
$29$ & $53$ & $35$ & $26$ & $25$ & $59$ & $ 4$ & $61$ & $54$ & $41$ & $ 9$ & $40$ & $ 8$ & $23$\\
\hline
$30$ & $24$ & $63$ & $31$ & $58$ & $45$ & $24$ & $62$ & $30$ & $11$ & $51$ & $16$ & $14$ & $40$\\
\hline
$31$ & $15$ & $23$ & $14$ & $47$ & $54$ & $46$ & $63$ & $31$ & $32$ & $ 1$ & $ 4$ & $24$ & $ 5$\\
\hline
\hline
\end{longtable}
\end{center}

As expected, since the degree is even $(n=6)$, the resulting composition is not unicyclic although the initial round corresponding to $\sigma_0$ is unicyclic.


\begin{thebibliography}{10}

\bibitem{BaBoHKTS2017}
\SortNoop{Bacher}{Bacher, Axel and Bodini, Olivier and Hwang, Hsien-Kuei and
  Tsai, Tsung-Hsi}.
\newblock Generating random permutations by coin tossing: Classical algorithms,
  new analysis, and modern implementation.
\newblock {\em ACM Trans. Algorithms}, 13(2):1--43, 2017.

\bibitem{Biham1991}
\SortNoop{Biham}{Biham, Eli and Shamir, Adi}.
\newblock Differential cryptanalysis of \uppercase{DES}-like cryptosystems.
\newblock {\em Journal of Cryptology}, 4(1):3--72, 1991.

\bibitem{BrKa1988}
\SortNoop{Brassard}{Brassard, Gilles and Kannan, Sampath}.
\newblock The generation of random permutations on the fly.
\newblock {\em Inf. Process. Lett.}, 28(4), July 1988.

\bibitem{Carlet1}
\SortNoop{Carlet}{Carlet, Claude}.
\newblock Boolean functions for cryptography and error correcting codes.
\newblock In Yves Crama and Peter L.~Hammer, editors, {\em Boolean Models and
  Methods in Mathematics, Computer Science, and Engineering}, pages 257--397.
  Cambridge University Press, 2010.
\newblock monography's chapter.

\bibitem{Carlet2}
\SortNoop{Carlet}{Carlet, Claude}.
\newblock Vectorial boolean functions for cryptography.
\newblock In Yves Crama and Peter L.~Hammer, editors, {\em Boolean Models and
  Methods in Mathematics, Computer Science, and Engineering}, pages 398--469.
  Cambridge University Press, 2010.
\newblock monography's chapter.

\bibitem{daemen2013design}
\SortNoop{Daemen}{Daemen, Joan and Rijmen, Vincent}.
\newblock {\em The design of Rijndael: AES-the advanced encryption standard}.
\newblock Springer Science \& Business Media, 2013.

\bibitem{Flajolet90randommapping}
\SortNoop{Flajolet}{Flajolet, Philippe and Odlyzko, Andrew M.}
\newblock Random mapping statistics.
\newblock In Jean-Jacques Quisquater and Joos Vandewalle, editors, {\em
  Advances in Cryptology --- EUROCRYPT '89}, volume 434 (LNCS), pages 329--354,
  1990.

\bibitem{flajolet2009analytic}
\SortNoop{Flajolet}{Flajolet, Philippe and Sedgewick, Robert}.
\newblock {\em Analytic Combinatorics}.
\newblock Cambridge University Press, 1st edition, 2009.

\bibitem{HM}
\SortNoop{Hansen}{Hansen, Tom and Mullen, Gary L}.
\newblock Primitive polynomials over finite fields.
\newblock {\em Mathematics of Computation}, 59(200):639--643, 1992.

\bibitem{XiaoM88}
\SortNoop{Massey}{{Xiao, Guo{-}Zhen and L.~Massey, James}}.
\newblock A spectral characterization of correlation-immune combining
  functions.
\newblock {\em {IEEE} Trans. Information Theory}, 34(3):569--571, 1988.

\bibitem{matsui1993linear}
\SortNoop{Matsui}{Matsui, Mitsuru}.
\newblock Linear cryptanalysis method for {DES} cipher.
\newblock In {\em Workshop on the Theory and Application of of Cryptographic
  Techniques}, pages 386--397. Springer, 1993.

\bibitem{MullenPanario2013}
\SortNoop{Mullen}{{Mullen, Gary L. and Panario, Daniel}}.
\newblock {\em Handbook of Finite Fields}.
\newblock Chapman \& Hall/CRC, 2013.

\bibitem{Nyberg2017}
\SortNoop{Nyberg}{Nyberg, Kaisa}.
\newblock Statistical and linear independence of binary random variables.
\newblock Cryptology ePrint Archive, Report 2017/432, 2017.

\bibitem{Sigenthaler1984}
\SortNoop{Siegenthaler}{Siegenthaler, T.}
\newblock Correlation-immunity of nonlinear combining functions for
  cryptographic applications.
\newblock {\em IEEE Trans. Info. Th.}, 30:776--780, 1984.

\bibitem{Szpankowski2001}
\SortNoop{Szpankowski}{Szpankowski, Wojciech}.
\newblock {\em Average Case Analysis of Algorithms on Sequences}.
\newblock John Wiley \& Sons, Inc., 2001.

\bibitem{Wan}
\SortNoop{Wan}{Wan, Daqing}.
\newblock Generators and irreducible polynomials over finite fields.
\newblock {\em Mathematics of Computation}, 66(219):1195--1212, 1997.

\end{thebibliography}
\end{document}